\documentclass[preprint,showpacs,nofootinbib]{revtex4}
\usepackage{amssymb}
\usepackage{dcolumn}
\usepackage{psfrag}
\usepackage{graphicx}

\def\be{\begin{equation}}
\def\ee{\end{equation}}
\def\bea{\begin{eqnarray}}
\def\eea{\end{eqnarray}}
\input epsf

\begin{document}
\begin{flushright} {\footnotesize  LMU-ASC 78/05}  
\end{flushright}

\title{{\LARGE Enhancing the tensor-to-scalar ratio in simple inflation}}
\author{V. Mukhanov$,$ A.Vikman}
\affiliation{ASC, Physics Department LMU, Theresienstr. 37, Munich, Germany}

\begin{abstract}
We show that in theories with a nontrivial kinetic term the contribution
of the gravitational waves to the CMB fluctuations can be substantially
larger than that is naively expected in simple inflationary models. This
increase of the tensor-to-scalar perturbation ratio leads to a larger
B-component of the CMB polarization, thus making the prospects for 
future detection much more promising. The other important consequence of the
considered model is a higher energy scale of inflation and hence higher
reheating temperature compared to a simple inflation.
\end{abstract}

\maketitle



\section{Introduction}
The main consequence of inflation is the generation of primordial
cosmological perturbations \cite{mc} and the production of long wavelength
gravitational waves (tensor perturbations)~\cite{star}. The predicted
slightly red-tilted spectrum of the scalar perturbations is at present in
excellent agreement with the measurements of the CMB fluctuations \cite{cmb}%
. The detection of a small deviation of the spectrum from flat
together with the observation of primordial gravitational waves would make
us completely confident in early-time cosmic acceleration. The
detection of primordial gravitational waves is not easy, but they can
be seen indirectly in the B-mode of the CMB polarization (see,
for example, \cite{book}). In standard slow-roll inflationary scenarios 
\cite{Chaot} the amplitude of the tensor perturbations can, in principle, be
large enough to be discovered. However, it is only on the border of detectability
in future experiments.

There is no problem to modify the inflationary scenarios in a way to
suppress the tensor component produced during inflation. In particular, in
models such as new inflation~\cite{New} and hybrid inflation~\cite{Hybrid},
tensor perturbations are typically small~\cite{book}. Moreover, in the
curvaton scenario \cite{curva} and k-inflation \cite{k-Inflation}, they can
be suppressed completely.

An interesting question is whether the gravitational waves can be
significantly enhanced compared to our naive expectations. Recently it was
argued that the contribution of tensor perturbations to the CMB anisotropy
can be much greater than expected \cite{Bartolo:2005jg,Sloth:2005yx}.
However, it was found in \cite{lms} that in the models considered in \cite%
{Bartolo:2005jg,Sloth:2005yx} one cannot avoid the production of too large
scalar perturbations and therefore they are in contradiction with
observations. It is also possible to produce blue-tilted spectrum of 
gravitational waves in a so-called superinflation \cite{baldi}, 
which is, however, plagued by graceful exit problem.    

At present there exists no inflationary model with graceful exit where the
B-mode of polarization would exceed that predicted by simple chaotic
inflation. The purpose of our paper is to show that such models can be
easily constructed even within the class of simple slow-roll inflationary
scenarios if we allow a nontrivial dependence of the Lagrangian on the
kinetic term. These models resemble k-inflation with only difference
that here inflation is due to the potential term in the Lagrangian.

\section{Basic equations}
The generic action describing a scalar field interacting with the
gravitational field is%
\begin{equation}
S=S_{g}+S_{\varphi }=\int d^{4}x\sqrt{-g}\left[ -\frac{R}{16\pi }+p(\phi ,X)%
\right] ,  \label{1}
\end{equation}%
where $R$ is the Ricci scalar and $p(\phi ,X)$ is a function of the scalar
field $\phi $ and its first derivatives 
\begin{equation}
X=\frac{1}{2}\nabla _{\mu }\phi \nabla ^{\mu }\phi .
\end{equation}%
We use Planck units, where $G=\hbar =c=1.$ In the case of the usual scalar
field the $X-$dependence of $p$ is trivial, namely, $p=X-V\left( \phi
\right) ,$ while k-inflation and k-essence \cite%
{k-Inflation,k-Essence,book} are based on the non-trivial dependence of $p$
on $X.$ For $X>0$, variation of the action (\ref{1}) with respect to the metric
gives the energy momentum tensor for the scalar field in the form of an
\textquotedblleft ideal hydrodynamical fluid\textquotedblright :\ 
\begin{equation}
T_{\nu }^{\mu }=(\varepsilon +p)u^{\mu }u_{\nu }-p\delta _{\nu }^{\mu }.
\label{2}
\end{equation}%
Here the Lagrangian $p(\phi ,X)$ plays the role of pressure, the energy
density is given by 
\begin{equation}
\varepsilon =2Xp_{,X}-p ,  \label{3}
\end{equation}%
where $p_{,X}=\partial p/\partial X,$ and the \textquotedblleft
four-velocity\textquotedblright\ is%
\begin{equation}
u_{\mu }=\frac{\nabla _{\mu }\phi }{\sqrt{2X}}.
\end{equation}
Let us consider a spatially flat Friedmann universe with small perturbations:
\begin{equation}
ds^{2}=\left( 1+2\Phi \right) dt^{2}-a^{2}(t)\,\left[ \left( 1-2\Phi \right)
\delta _{ik}+h_{ik}\right] dx^{i}dx^{k},  \label{4}
\end{equation}%
where $\Phi $ is the gravitational potential characterizing scalar metric
perturbations and $h_{ik}$ is a traceless, transverse perturbations
describing the gravitational waves. The scale factor $a\left( t\right) $
satisfies the Friedmann equation%
\begin{equation}
H^{2}\equiv \left( \frac{\dot{a}}{a}\right) ^{2}=\frac{8\pi }{3}\varepsilon ,
\label{fe}
\end{equation}%
where the dot denotes the derivative with respect to time $t.$ Taking into
account that for the homogeneous background scalar field $X=\frac{1}{2}\dot{%
\phi}^{2}$ and that the energy conservation law is
\begin{equation}
\dot{\varepsilon}=-3H\left( \varepsilon +p\right) ,  \label{ec}
\end{equation}%
we obtain the following equation for the scalar field%
\begin{equation}
\ddot{\phi}+3c_{S}^{2}H\dot{\phi}+\frac{\varepsilon _{,\phi }}{\varepsilon
_{,X}}=0,  \label{esf}
\end{equation}%
where the \textquotedblleft speed of sound\textquotedblright\ is 
\begin{equation}
c_{S}^{2}\equiv \frac{p_{,X}}{\varepsilon _{,X}}=\left[ 1+2X\frac{p_{,XX}}{%
p_{,X}}\right] ^{-1} . \label{ss}
\end{equation}%
Considering small scalar metric perturbations one can show
that $c_{S}$ is in fact the speed of propagation of the cosmological
perturbations \cite{mg,book}. The stability condition with respect to the
high frequency cosmological perturbations requires $c_{S}^{2}>0$.

\section{Generalized slow-roll inflation}
It follows from (\ref{3}) that inflation can be realized if the condition $%
Xp_{,X}\ll p$ is satisfied for a sufficiently long time interval. This can be
done in two ways. Considering the canonical scalar field with $p=X-V\left(
\phi \right) $, one can take a flat potential $V\left( \phi \right) $ so that 
$X\ll V$ for more than 75 e-folds. This is the standard slow-roll inflation 
\cite{Chaot} and in this case $c_{S}=1.$ The other possibility is the
k-inflation \cite{k-Inflation}, where $p$ is a weakly dependent function of $%
X$, so that $p_{,X}$ is small. Here inflation is entirely based on the
kinetic term and it can take place even if the field is running very fast ($X
$ is large); typically $c_{S}^{2}\ll 1$ for k-inflation.

In this paper we consider a slow-roll inflation with a flat potential but in
theories with a nontrivial kinetic term. To best of our knowledge this
possibility has been ignored in the literature until now. 
However such models are
very interesting because, as we will see, they allow us to have $c_{S}^{2}>1$
during inflation and thus enhance the tensor-to-scalar ratio for the
perturbations produced. 

The Lagrangian $p(\phi,X)$ is manifestly Lorentz-invariant and 
a superluminal speed of cosmological perturbations  
does not contradicts the principles of relativity.
In fact, the superluminal propagation is possible 
only in the presence of the time-dependent homogeneous 
scalar field, which determines a preferable coordinate frame.
Only in this frame the speed can exceed the speed of light.
As a result no violation of causality is possible.  

For simplicity let us consider theories with
Lagrangian 
\begin{equation}
p=K(X)-V(\phi ).
\end{equation}%
In this case 
\begin{equation}
\varepsilon =2XK_{,X}-K+V,  \label{5}
\end{equation}%
and the equation for scalar field becomes 
\begin{equation}
\ddot{\phi}+3c_{S}^{2}H\dot{\phi}+\frac{V_{,\phi }}{\varepsilon _{,X}}=0.
\label{6}
\end{equation}%
It is clear that if the slow-roll conditions 
\[
XK_{,X}\ll V,\text{ }K\ll V,\text{ }\left\vert \ddot{\phi}\right\vert \ll 
\frac{V_{,\phi }}{\varepsilon _{,X}}
\]%
are satisfied for at least 75 e-folds then we have a successful slow-roll
inflation due to the potential $V.$ In contrast to ordinary
slow-roll inflation one can arrange here practically any speed of sound $%
c_{S}^{2}$ by taking an appropriate kinetic term. For example, for%
\[
K\left( X\right) =\alpha X^{\beta },
\]%
we obtain from (\ref{ss})%
\[
c_{S}^{2}=1/\left( 2\beta -1\right) .
\]%
and if $\beta \rightarrow 1/2$, then $c_{S}^{2}\rightarrow \infty .$ Therefore
by considering nontrivial kinetic terms $K\left( X\right) $ one
can have, in principle, an arbitrarily large  $c_{S},$ which thus becomes an
additional free parameter of the theory. The crucial point is that 
the amplitude of the final
scalar perturbations (during the postinflationary, radiation-dominated
epoch) depends on $c_{S}$ (see, \cite{book}): 
\begin{equation}
\delta _{\Phi }^{2}\simeq \frac{64}{81}\left( \frac{\varepsilon }{%
c_{S}\left( 1+p/\varepsilon \right) }\right) _{c_{S}k\simeq Ha},  \label{scp}
\end{equation}%
while the ratio of tensor to scalar amplitudes on supercurvature scales is
given by 
\begin{equation}
\frac{\delta _{h}^{2}}{\delta _{\Phi }^{2}}\simeq 27\left( c_{S}\left( 1+%
\frac{p}{\varepsilon }\right) \right) _{k\simeq Ha}.  \label{tsr}
\end{equation}%
Here it is worthwhile reminding that all physical quantities on the right hand 
side of Eqs. (\ref{scp}) and (\ref{tsr}) have to be calculated during inflation 
at the moment when perturbations with 
wave number $k$ cross corresponding Horizon: 
$c_{S}k\simeq Ha$ for (\ref{scp}) and  $k\simeq Ha$ for (\ref{tsr}) respectively.   
The amplitude of the scalar perturbations $%
\delta _{\Phi }$ is a free parameter of the theory which is taken to fit the
observations. Therefore, in models where $c_{S}^{2}>1$, the energy scale
of inflation must be higher than in ordinary slow-roll inflation with
canonical kinetic term. Moreover, it follows from (\ref{tsr}) that the
tensor-to-scalar ratio can be arbitrarily enhanced in such models.

\section{Simple model}
As a concrete example from the class of theories with enhanced tensor
contribution let us consider a simple model with Lagrangian 
\begin{equation}
p(\phi ,X)=\alpha ^{2}\left[ \sqrt{1+\frac{2X}{\alpha ^{2}}}-1\right] -\frac{%
1}{2}m^{2}\phi ^{2},  \label{sm1}
\end{equation}%
where constant $\alpha $ is a free parameter. For $2X\ll \alpha ^{2}$ one
recovers the Lagrangian for the usual free scalar field. The function $p$ is
a monotonically growing concave function of $X,$%
\[
p_{,X}=\left( 1+\frac{2X}{\alpha ^{2}}\right) ^{-1/2}>0,\text{ \ \ \ \ }%
p_{,XX}=-\frac{1}{\alpha ^{2}}\left( 1+\frac{2X}{\alpha ^{2}}\right)
^{-3/2}<0,
\]%
and the corresponding energy density, 
\begin{equation}
\varepsilon =\alpha ^{2}\left[ 1-\left( 1+\frac{2X}{\alpha ^{2}}\right)
^{-1/2}\right] +\frac{1}{2}m^{2}\phi ^{2},  \label{sm2}
\end{equation}%
is always positive. The effective speed of sound,%
\begin{equation}
c_{S}^{2}=\frac{p_{,X}}{\varepsilon _{,X}}=1+\frac{2X}{\alpha ^{2}},
\label{sm4}
\end{equation}%
is larger than the speed of light, approaching it as $X\rightarrow 0.$

\begin{figure}[t]
\begin{center}
\psfrag{fd}[bl]{$\dot\phi$}
\psfrag{f}[bl]{$\phi$}
\psfrag{A}[bl]{\small$attractor$}
\psfrag{fe}[br]{$\phi_{end}$}
\psfrag{Fa}[l]{$\frac{mc_{\star}}{\sqrt{12\pi }}$}
\psfrag{FaN}[br]{$-\frac{mc_{\star}}{\sqrt{12\pi }}$}

\includegraphics[width=\textwidth]{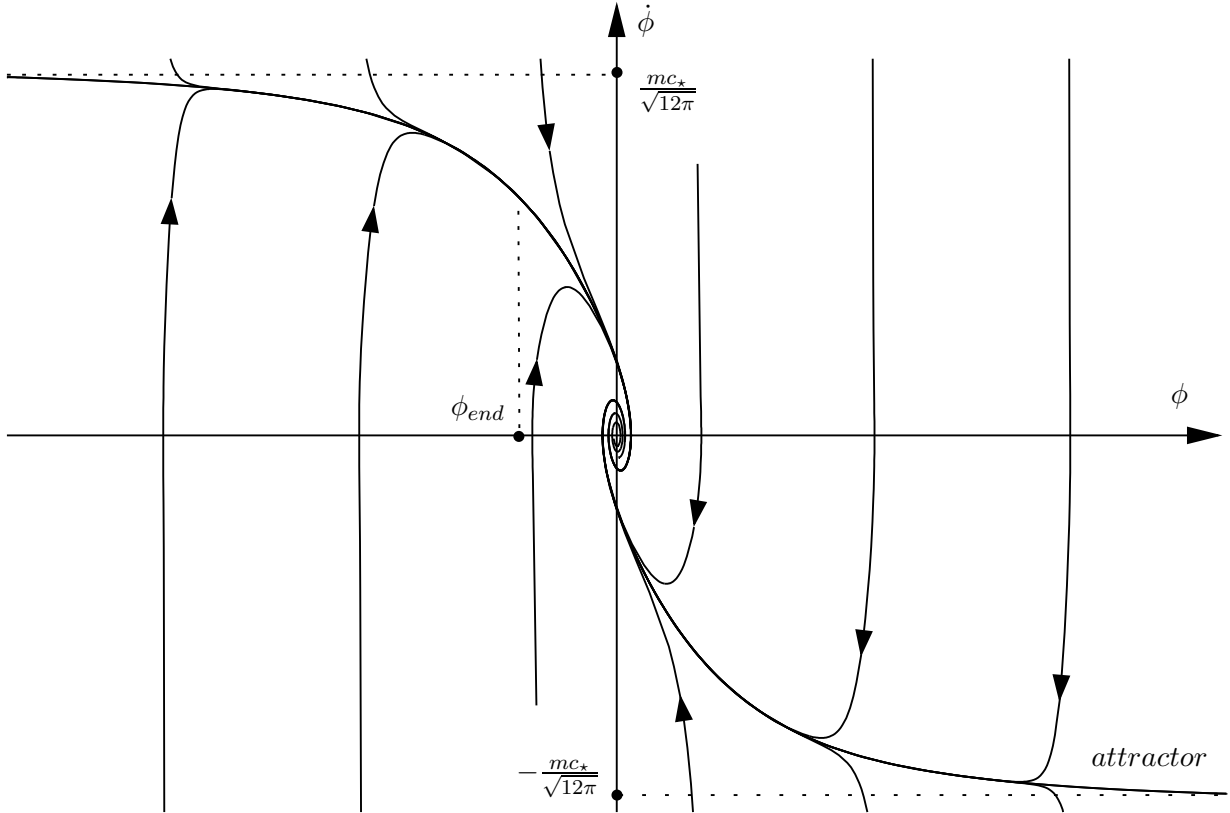}
\caption{\label{phase}Numerically obtained phase portrait for the system with
the parameters $m=1.5 \cdot 10^{- 7}$, $c_{\star}=3.67$. 
The field value $\phi_{end}=0.32$ 
corresponds to the end of acceleration when $w=-1/3$.}
\end{center}
\end{figure}

In the slow-roll regime and for $p$ given in (\ref{sm1}), equations (\ref{fe}), (%
\ref{esf}) simplify to%
\begin{equation}
H\simeq \sqrt{\frac{4\pi }{3}}m\phi ,\text{ \ \ \ }3p_{,X}H\dot{\phi}%
+m^{2}\phi \simeq 0.  \label{sm5}
\end{equation}%
Taking into account that $\dot{\phi}=-\sqrt{2X}$, we infer that during
inflation $\dot{\phi}$ is constant and 
\begin{equation}
\frac{2X}{\alpha ^{2}}=\left( \frac{12\pi \alpha ^{2}}{m^{2}}-1\right) ^{-1}.
\label{sm6}
\end{equation}%
The speed of sound 
\begin{equation}
c_{S}\simeq c_{\star }=\left( 1-\frac{m^{2}}{12\pi \alpha ^{2}}\right) ^{-1/2}
\label{sm7}
\end{equation}%
can be arbitrarily large during inflation if we take $12\pi \alpha
^{2}\rightarrow m^{2}$. Note, however, that $12\pi\alpha^{2}>m^{2}$ 
is necessary for the existence of the slow-roll solution.
Hereafter we will use $c_{\star}$ as a parameter
instead of $\alpha$. One can easily find that during the slow-roll regime 
\begin{equation}
\dot{\phi}\simeq -\frac{mc_{\star}}{\sqrt{12\pi }},  \label{sm8}
\end{equation}%
and the pressure and energy density are given by 
\begin{equation}
p\simeq m^{2}\left( \frac{1}{12\pi }\frac{c_{\star }^{2}}{1+c_{\star }}-%
\frac{\phi ^{2}}{2}\right) ,\text{ \ }\varepsilon \simeq m^{2}\left( \frac{1%
}{12\pi }\frac{c_{\star }}{1+c_{\star }}+\frac{\phi ^{2}}{2}\text{\ }\right)
,  \label{sm9}
\end{equation}%
respectively. The inflation is over when 
\begin{equation}
\frac{\varepsilon +p}{\varepsilon }\simeq \frac{c_{\star}}{6\pi \phi ^{2}}
\label{sm10}
\end{equation}%
becomes of order unity, that is, at $\phi \sim \sqrt{c_{\star}/6\pi }.$
After that the field $\phi $ begins to oscillate and decays.
The detailed analysis of dynamics and stability of the model will be given 
in \cite{mv}. Here we restrict ourselves to illustration of the above analysis 
by a phase diagram Fig.~\ref{phase} obtained numerically. 

To determine $a\left( \phi \right) $ we use (\ref{sm8}) to rewrite the first
equation in (\ref{sm5}) as%
\begin{equation}
-\frac{mc_{\star }}{\sqrt{12\pi }}\frac{d\ln a}{d\phi }\simeq \sqrt{\frac{%
4\pi }{3}}m\phi ,  \label{sm11}
\end{equation}%
and obtain%
\begin{equation}
a\left( \phi \right) \simeq a_{f}\exp \left( \frac{2\pi }{c_{\star }}\left(
\phi _{f}^{2}-\phi ^{2}\right) \right) ,  \label{sm12}
\end{equation}%
where $a_{f}$ and $\phi _{f}\sim \sqrt{c_{\star}/6\pi }$ are the values of
the scale factor and the scalar field at the end of inflation. Given a number of
e-folds before the end of inflation $N,$ we find that at this time 
\begin{equation}
\frac{2\pi \phi ^{2}}{c_{\star}}\simeq N,
\end{equation}%
and, hence, 
\[
\frac{\varepsilon +p}{\varepsilon }\simeq \frac{1}{3N}
\]%
does not depend on $c_{\star}.$ Thus, for a given scale, which crosses the
Hubble scale $N$ e-folds before the end of inflation, the tensor-to-scalar
ratio is 
\begin{equation}
\frac{\delta _{h}^{2}}{\delta _{\Phi }^{2}}\simeq 27c_{\star }\left( 1+\frac{%
p}{\varepsilon }\right) \simeq \frac{9c_{\star}}{N}.
\end{equation}%
It is clear that by choosing $\alpha $ close to the critical value $m/\sqrt{%
12\pi }$ we can have a very large $c_{\star}$ and consequently enhance this
ratio almost arbitrarily.

\section{Conclusions}

We have shown above that in theories where the Lagrangian is a
nontrivial, nonlinear function of the kinetic term, the scale of inflation
can be pushed to a very high energies without coming into conflict 
with observations.
As a result, the amount of produced gravitational waves can be much
larger than is usually expected. If such a situation were realized in
nature then the prospects for the future detection of the B-mode of CMB
polarization are greatly improved.  
Of course, the theories where this happens
are somewhat \textquotedblleft fine-tuned\textquotedblright . Namely, the
corresponding Lagrangian $p\left( X,\phi \right) $ must generically satisfy
the condition, $p_{,X}+2Xp_{,XX}\ll p_{,X},$ during inflation. For the 
particular
model (\ref{sm1}) this means that $12\pi \alpha ^{2}-m^{2}\ll12\pi \alpha ^{2}$. This
fine-tuning of the parameters of the theory should not be confused, however,
with the fine-tuning of the initial conditions. In fact, the true theory of
nature is unique and, for example, Lagrangian (\ref{sm1}), where $12\pi
\alpha ^{2}$ is not very different from $m^{2}$ looks even more attractive
because it has fewer free parameters. Therefore, future observations of
the CMB fluctuations are extremely important since they will restrict
the number of possible candidates for the inflaton. The other peculiar
properties of the generalized slow-roll inflation will be considered in 
\cite{mv}.

\section*{Acknowledgements}
We are grateful to Matthew Parry for very useful discussions 
and  Marco Baldi, Fabio Finelli, Sabino Matarrese for useful correspondence.

\end{document}